\documentclass[article,preprintnumbers,amsmath,amssymb]{revtex4}

\usepackage{graphicx}
\usepackage{dcolumn}
\usepackage{bm}

\begin{document}

\preprint{In: Proc. of Int. Conf. ``New Trends in High-Energy
Physics", Sep.27 -- Oct.4, 2008, Yalta, Crimea}

\title{Issues of Reggeization in $qq'$ Back-Angle Scattering}
\author{M. V. Bondarenco\\
\small \it  NSC Kharkov Institute of Physics \& Technology,\\ 1
Academicheskaya St., Kharkov 61108, Ukraine}
%
%
\begin{abstract}
The Kirschner-Lipatov result for the DLLA of high-energy $qq'$
backward scattering is re-derived without the use of integral
equations. It is shown that part of the inequalities between the
variables in the logarithmically-divergent integrals is
inconsequential. The light-cone wave-function interpretation under
the conditions of backward scattering is discussed. It is argued
that for hadron-hadron scattering in the valence-quark model the
reggeization should manifest itself at full strength starting from
$s_{hh}=50~GeV^2$.
\end{abstract}

\maketitle

\section{Introduction}

Backscattering of high and intermediate energy, weakly
radiating\footnote{Weakness of the radiation is, rather, a wish then
a necessary condition. Backscattering, or large-angle scattering of
electrons, of course, is widely applied, too, but requires proper
calculation of radiative corrections.} particles (protons, X-ray) is
known as a clean tool for atomic material structure analysis
\cite{Feldman}. The clarity of the analysis owes exactly to the low
scattered particle fraction. With the initial macroscopic
luminosity, that poses no problem for detectability, but, more
importantly, the relative background (actually, all the diverse
kinds thereof) is suppressed.

In case when wave nature of the scattered particles is relevant, the
backward scattering can sometimes get enhanced as compared with that
at other large angles -- due to some or the other kinematical
symmetry (coherent backscattering).

For hadrons, which are objects composed of quarks, one can detect
the events of single quark backscattering by separating (single-)
flavor exchange reactions to high energy. Thereat, most probably,
only one pair of quarks (of different flavor) scatters backwards and
then recombines with the forward-moving hadron remnants. There is no
external gluonic radiation in the fully exclusive reaction, because
of color confinement. Besides, there is no necessity to rise the
energy to extremely high values, where some internal radiative
effects should eventually become important.

Owing to the hardness present in the process, a plausible
approximation for it is one-gluon exchange. The latter is impact
parameter conserving, which is convenient for the overlap
representation of the scattering matrix element in terms of quark
wave functions of hadrons, as was partially discussed elsewhere
\cite{Bond-Dubna}.

But yet, at energies high enough, the energy dependence of
flavor-exchange reactions departs notably from the
one-gluon-exchange prediction $\sigma\sim s^{-2}$, which is referred
to as reggeization phenomenon. It is desirable to get it
incorporated in the theory, within the impulse approximation
treatment.

In 1967, Gorshkov, Gribov, Lipatov, and Frolov \cite{FGGL} (see also
textbook \cite{BLP}) had evaluated double-leading-logarithmic
asymptotics (DLLA) of Feynman integrals corresponding to $e^-\mu^-$
back-angle scattering in QED, and resumed to all orders. They had
found a power falloff slowdown (basically, $t$-independent).

Later, Kirschner \cite{Kirschner}, being generally interested in
DLLA of QCD elementary scattering processes, examined quark-quark
backward scattering, and quark-antiquark forward/backward
annihilation, paralleling the framework of \cite{FGGL}. The negative
signature amplitude was thereafter computed by Kirschner and Lipatov
\cite{KL}.

The amplitude of $qq'$ backward scattering, which is the kernel of
the hadron binary reaction overlap matrix element, has the
asymptotics
\begin{eqnarray}
    M_{qq'\to
    q'q}\left(s_{qq'}\right)&=&\frac1{N_c}\delta_{m'l}\delta_{l'm}\sqrt{\frac{2\pi\alpha_s}{C_F}}\frac{8\pi}{\ln s_{qq'}}I_1\left(\sqrt{\frac{2\alpha_sC_F}{\pi}}\ln s_{qq'}\right)\quad (\texttt{DLLA})\label{regge th}\\
    &\sim&\frac{1}{\ln s_{qq'}}s_{qq'}^{\sqrt{\frac{2\alpha_sC_F}{\pi}}}\qquad\qquad\qquad\qquad\qquad\qquad\qquad\quad
    (s\to\infty)\label{regge asympt}
\end{eqnarray}
with $I_1$ the modified Bessel function, and
\begin{equation}\label{C_F}
    C_F=\frac{N_c^2-1}{2N_c}.
\end{equation}
(Account of single logarithms can somewhat change the index in
(\ref{regge th}), but the effect of that correction is rather
uncertain in view of our poor knowledge of the coupling constant
$\alpha_s$, anyway.)

Letting numerically $N_c=3$, $\alpha_s\simeq0.1\div0.2$, and
assuming that for reactions such as $np\to pn$ small Feynman-$x$
contribution is moderate (given that constituent quark models work
rather well for nucleon), one obtains an estimate
\begin{equation}\label{power-num}
    \frac{d\sigma_{np\to pn}}{dt}\propto\frac{1}{s^2}\left|M_{du\to ud}\right|^2\sim s^{-1.4\div-1.2}.
\end{equation}
This asymptotics is expected to hold when
$\sqrt{\frac{2\alpha_sC_F}{\pi}}\ln\left(s_{qq'}=\frac{s_{hh}}{N_1N_2}\right)\gg1$,
where $N_1$, $N_2$ are valence quark numbers in the colliding
hadrons. That numerically implies
\begin{equation}\label{s>>}
    s_{hh}\gg50\div100~{\rm \texttt{\textrm{GeV}}}^2.
\end{equation}

The correspondence of (\ref{power-num}) with the experimental
behavior is \emph{not} too bad.

The best experimental representative of flavor exchange reactions is
$np\to pn$, given the detailed data available for $d\sigma/dt$ and
even some data for polarization for this reaction, and in addition
-- nucleon form-factors as an independent constraint for the wave
function.

\begin{figure}[b]
\includegraphics[scale=.65]{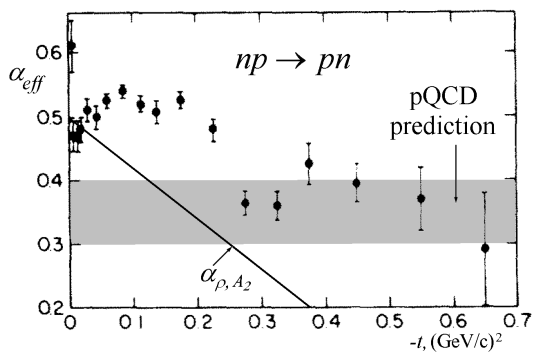}
%
%
\caption{Regge trajectory for $np\to pn$ reaction. Data taken from
\cite{Barton}. The straight line shown for comparison is the
conventional $\rho, A_2$-trajectory $0.5+0.8t.$}
\label{fig:trajectory}       
\end{figure}

The Regge trajectory slope for $np\to pn$ is small (see
Fig.~\ref{fig:trajectory}). In contrast, for meson flavor exchange,
particularly for $\pi^-p\to\pi^0n$ (usually quoted as having an
examplary linear Regge trajectory) the slope seems to be close to
Chew-Frautchi substantial value 0.8~\textrm{GeV}$^{-2}$. But it is
to be minded that in the pion charge exchange case there are
cancelations between $ud\to du$ scattering and $u\bar u\to d\bar d$
annihilation, and in itself, pion is a more relativistic system then
nucleon, probably, with a larger contribution from small $x$.
Altogether, this makes the dynamics more intricate, and we refrain
from discussing it here.

%

In this contribution we shall focus only on reggeization of
two-free-quark scattering. That was the subject of Kirschner and
Lipatov, but it is desirable to give it more dynamical
interpretation, which can in future prove useful for scattering
treatment in the spectator quark surroundings.

\section{The origin of enhancements}\label{sec:enh origin}
Consider an ultra-relativistic collision of a free $d$-quark
carrying momentum $p_d$ with a free $u$-quark of momentum $p_u$,
resulting in a near-backward elastic scattering to momenta $p'_d$,
$p'_u$:
\[
    d(p_d)+u(p_u)\to u(p'_u)+d(p'_d),
\]
\[
\Delta_\perp=p_d-p'_d=p'_u-p_u\sim1~{\rm \texttt{\textrm{GeV}}}.
\]
As long as no other particles are concerned in the initial or final
state, we shall throughout designate inter-quark kinematic
invariants without hats or subscripts:
\[s=(p_d+p_u)^2\gg\Delta_\perp^2.\]
Quark bispinors will be denoted as $u$ for initial and $u'$ for the
final $u$-quark, and $d$, $d'$ -- the same for $d$-quark.

\subparagraph{\textbf{Loop structure. Collinear vs. infra-red large
logarithms}} The tree-level amplitude of quark-quark back-angle
scattering is the single-gluon exchange:
\begin{equation}\label{}
    M^{(1)}\approx-\frac{4\pi\alpha}{s}\left(\bar d'\gamma^\mu d\right)\left(\bar u'\gamma^\mu
    u\right)t^A_{l'l}t^A_{m'm}.
\end{equation}
It scales with the collision energy $\sqrt s$ as $s^0$, which
corresponds to cross-section decreasing as $s^{-2}$. But in higher
orders there can arise loop enhancements of logarithmic kind, which
are conventionally classified by two categories -- soft and
collinear ones. Soft divergences originate when some mass ratio
tends to be large, $\frac{m}{\lambda}\to\infty$. Collinear ones
require the high-energy limit $\frac{s}{m^2}\to\infty$; they
correspond to an effective phase space extension with the
energy\footnote{An often quoted definition of collinear divergence
type is that it is inherent to the massless case, when the
singularity of the integrand is encountered not in a single point
(that would characterize soft divergence), but along an entire line.
But from the viewpoint of initially massive, physical case, one yet
needs to specify, in which order the massless limit is achieved. The
answer is that ratios of all the masses stay finite and non-zero,
while their ratios to the energy tend to zero -- in contrast to the
soft case when some mass ratio turns small. Again, that is
equivalent to the growth of the effective phase space, in mass
units. As for the method of identifying soft and collinear
divergences by $\frac{d\omega}{\omega}$ and $\frac{d\theta}{\theta}$
factors, it is not manifestly Lorentz-invariant.}.

In general, a collinear divergence is encountered when a soft
virtual particle connects two high-energy lines, provided the latter
are sufficiently close to the mass-shell. Then, the high-energy line
propagators admit eikonal approximation\footnote{The eikonal
condition ($pk\gg k^2,p^2-m^2$) is exactly the criterion of the line
proximity to the mass shell.}, $\sim\frac{1}{pk}$ ($p$ being the
momentum of the high-energy line, and $k$ -- the momentum of the
soft one), whereas the soft particle propagator decreases as
$\sim\frac{1}{k^2}$ if it is a boson, or $\sim\frac{\not k}{k^2}$ if
it is a fermion. When covered with 4d integration, by $k$-power
counting it is seen to produce logarithmic divergences -- in a
triangle loop with two eikonal (fermion) and one soft boson lines
(not counting possible hard propagators, which may be regarded as
momentum-independent, and graphically represented as contracted into
a point), and in quadrangular loops with two eikonal (boson) lines
and two soft fermion lines.

In the first case, of triangular loops, the collinear divergence is
merging with the soft one (IR). Although those can be given
independent meaning, physically they both are related to emission
and reabsorption of bremsstrahlung photons, with the energy smaller
then the mass of the radiating particle (in the IR soft case), or
then the collision energy (in the IR collinear case). So, it is
natural that they obey the same cancelation principles. For
back-angle scattering of equal-charge particles, or with perfect
charge (color) exchange, IR cancelations must be working at full
strength.

The second case, of quadrangular loop, instead, has no soft
counterpart. Moreover, virtual corrections of that kind upon
resummation should lead to enhancement rather then suppression of
the cross-section, as we shall discuss in detail below.

In higher orders of perturbation theory, in order to obtain the
leading logarithmic contribution, there must be an eikonal condition
for each gluon line. Denoting by $q_i$ -- $d$-quark momenta on its
course from $p_d$ to $p'_d$ (see Fig.~\ref{fig:diagrams} below),
\begin{equation}\label{eikonal i}
    -(q_{i-1}-q_i)^2\approx2q_{i-1}q_i\gg q_{i-1}^2,q_i^2.
\end{equation}
Fulfilment of these conditions is possible if the intermediate quark
(and gluon) momenta approximately belong to the plane formed by
initial and final momenta. For back-angle scattering this plane
approximately coincides with that of collision, and it may
unequivocally be called longitudinal. It is profitable to define in
it light-cone coordinates, and expand any vector
$a^\mu=a_\parallel^\mu+a_\perp^\mu$, $a_\parallel^\mu=(a^0, a^3)$,
$a_\perp^\mu=(a^1, a^2)$
\[
a^\pm=\frac{a^0\pm a^3}{\sqrt2},
\]
\[
a\cdot b=a^+b^-+a^-b^++a_\perp\cdot b_\perp.
\]
Then, Eq.~(\ref{eikonal i}) requires\footnote{Presuming that there
are no fine cancelations between $\parallel$ and $\perp$ components
in momentum squares, which would reduce the integration volume.}
\begin{equation}\label{+ ordering}
    p_d^+\gg q_1^+\gg q_2^+\ldots\gg q_{n-1}^+\gg p'^+_d,
\end{equation}
\begin{equation}\label{}
    p_d^-\ll q_1^-\ll q_2^-\ldots\ll q_{n-1}^-\ll p'^-_d,
\end{equation}
\begin{equation}\label{q perp ll q+q-}
    q_{i\perp}^2\ll q_{i-1}^+q_i^-,\, q_i^+q_{i+1}^-.
\end{equation}
In fact, Eq.~(\ref{q perp ll q+q-}) will be satisfied automatically
if
\begin{equation}\label{q_i^2>0}
    q_{i\perp}^2<2q_i^+q_i^-=q_{i\parallel}^2,\qquad q_i^2>0.
\end{equation}
This is nothing but the usual multi-peripheral kinematics -- the
same as for the reggeization at forward scattering. That is quite
natural, since the denominator structure for those cases is the same
(for instance, in a scalar theory, with no propagator numerators,
and all the particles identical, there would be no difference
between forward and backward scattering).

\begin{figure}[b]
\includegraphics[scale=.65]{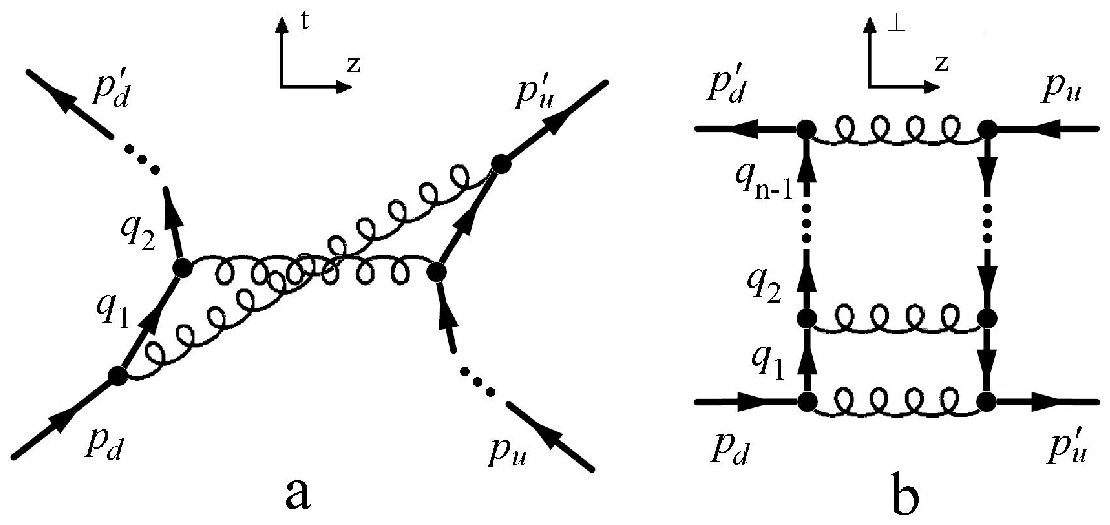}
%
%
\caption{The diagram giving leading logarithmic contribution in
$2n$-th order:\newline \textrm{a} -- the temporal ordering
representation;\newline \textrm{b} -- the spatial projection.}
\label{fig:diagrams}       
\end{figure}

\subparagraph{\textbf{The Feynman diagram topology.}} The ordering
in rapidity guarantees uniqueness of the Feynman diagram, and the
temporal order of boson emission from one fermion should be reverse
to that of their absorption by the other fermion. Thereby, the
concept of near-neighbor interaction in the phase space finds
support.

The amplitude corresponding to the $n$-rung diagram (see
Fig.~\ref{fig:diagrams}) is
\begin{equation}\label{}
    M^{(n)}=i(-4\pi
    i\alpha_s)^n \mathsf
    C^{(n)}\int\frac{d^4q_1}{(2\pi)^4}\ldots\frac{d^4q_{n-1}}{(2\pi)^4}\frac{\mathcal N^{(n)}}{\mathcal D^{(n)}}
\end{equation}
with
\begin{equation}\label{Cn}
    \mathsf
    C^{(n)}=\left(t^{A_n}\ldots t^{A_1}\right)_{l'l}\left(t^{A_1}\ldots
    t^{A_n}\right)_{m'm},
\end{equation}
\begin{equation}\label{Nn}
    \mathcal N^{(n)}\approx\left[\bar d'\gamma^{\mu_n}\not q_{n-1}\ldots\gamma^{\mu_2}\not q_1\gamma^{\mu_1}d\right]\left[\bar u'\gamma^{\mu_1}\not q_1\gamma^{\mu_2}\ldots\not
    q_{n-1}\gamma^{\mu_n}u\right],
\end{equation}
\begin{equation}\label{Dn}
    \mathcal D^{(n)}\approx(-2p_dq_1+i0)(-2q_1q_2+i0)(-2q_{n-1}p'_d+i0)q_1^2(q_1-\Delta)^2\ldots
    q_n^2(q_n-\Delta)^2.
\end{equation}

At this stage, certain insight can already be gained from the
topology of Feynman diagrams. Drawing Feynman diagrams in accord
with the process spatial projection, when the initial particle
momentum directions are opposite, an arbitrary order diagram is
depicted as a ladder, each cell of which is dissectible by two lines
in the $t$-channel (for backward scattering, $\left|t\right|\ll
s\approx\left|u\right|$). On the other hand, drawing Feynman
diagrams according to the event temporal ordering, either the 2
fermion lines, or all the boson lines must cross. In any way, the
diagram cannot be cut by two lines in the $s$-channel (which might
be utilized for evaluation by unitarity). This is in contrast with
the IR boson attachment order, where triangle loops (though not
necessarily the entire diagram) can always be cut by two lines in
the $s$-channel, and the concept of correspondence with the emitted
real bosons through unitarity is useful.

\subparagraph{\textbf{Classical interpretation.}} In classical
terms, the mechanism of enhancement may be thought of, roughly, as
follows. In a high-energy collision, charged particles can shed
their proper fields with the impart to them of the bulk of their
energy, and slow down. In the slow state, they are turned around on
a larger mutual distance, which results in the increase of the
scattering differential cross-section. Upon the reflection, the
charged particles can again pick up each the comoving proper field
from the other particle, and thus restore the high relative energy
up to the initial value.

\section{Numerators}\label{sec:numerators}
Let us, in the first place, analyze matrix numerators, determining
all the speciality of quark-quark scattering.
\subparagraph{\textbf{Spin factors.}}\label{sec:spin} As long as
fermion masses are neglected, their helicity must be conserved. But,
in addition, we shall acquire strict correlation of helicities of
colliding particles.

In addition to the light-cone decomposition, it is convenient to
introduce chiral vector basis in the transverse plane:
\begin{equation}\label{}
    a^R=-\frac{a^1+ia^2}{\sqrt2}, \qquad
    a^L=\frac{a^1-ia^2}{\sqrt2}, \qquad a_\perp\cdot b_\perp=a^Rb^L+a^Lb^R.
\end{equation}
Using in capacity of basic $\gamma$-matrices
\begin{equation}\label{}
    \gamma^\pm=\frac{\gamma^0\pm\gamma^3}{\sqrt2}, \qquad
    \gamma^R=-\frac{\gamma^1+i\gamma^2}{\sqrt2}, \qquad
    \gamma^L=\frac{\gamma^1-i\gamma^2}{\sqrt2}
\end{equation}
makes the covariant anticommutation relation
$\left\{\gamma^\mu,\gamma^\nu\right\}=2g^{\mu\nu}$ look like
\begin{equation}\label{}
    \left\{\gamma^+,\gamma^-\right\}=2, \qquad
    \left\{\gamma^R,\gamma^L\right\}=2,
\end{equation}
with all other anticommutators zero:
\begin{equation}\label{gamma squares}
    \left(\gamma^+\right)^2=\left(\gamma^-\right)^2=\left(\gamma^R\right)^2=\left(\gamma^L\right)^2=0,
\end{equation}
\begin{equation}\label{}
    \left\{\gamma^\pm,\gamma^{R,L}\right\}=0.
\end{equation}

Important for the future practice are cubic relations
\begin{equation}\label{}
    \gamma^R\gamma^L\gamma^R=2\gamma^R,\qquad
    \gamma^L\gamma^R\gamma^L=2\gamma^L,
\end{equation}
and Dirac conjugation properties
\begin{equation}\label{}
    \bar{\gamma^\pm}=\gamma^\pm,\qquad \bar\gamma^R=-\gamma^L, \qquad
    \bar\gamma^L=-\gamma^R.
\end{equation}

For in- and out- quark bispinors, which satisfy (massless) Dirac
equations
\begin{equation}\label{massless Dirac}
    \gamma^-d=0,\qquad \bar u'\gamma^-=0,\qquad \bar d'\gamma^+=0,\qquad
    \gamma^+u=0,
\end{equation}
further, define polarization states as those of definite helicity
(left and right):
\begin{equation}\label{}
    \gamma^Ld_R=\sqrt2d_L,\qquad \gamma^Rd_L=\sqrt2d_R,
\end{equation}
\begin{equation}\label{r1}
    \bar u'_R\gamma^R=-\overline{\gamma^Lu'_R}=-\sqrt2\bar u'_L, \qquad
    \bar u'_L\gamma^L=-\sqrt2\bar u'_R,
\end{equation}
\begin{equation}\label{}
    \gamma^Ru_R=-\sqrt2u_L,\qquad \gamma^Lu_L=-\sqrt2u_R,
\end{equation}
\begin{equation}\label{}
    \bar u'_R\gamma^L=\sqrt2\bar d'_L, \qquad \bar
    d'_L\gamma^R=\sqrt2\bar d'_R,
\end{equation}
and the normalization should be
\begin{equation}\label{r4}
    \bar d'_Ld_R=\bar d'_Rd_L=\bar u'_Lu_R=\bar u'_Ru_L=\sqrt s.
\end{equation}

The important consequence of Eqs.~(\ref{r1}-\ref{r4}) and
(\ref{gamma squares}) is
\begin{equation}\label{}
    \gamma^Rd_R=0,\qquad \gamma^Ld_L=0.
\end{equation}
\begin{equation}\label{}
    \bar u'_R\gamma^L=0,\qquad \bar u'\gamma^R=0.
\end{equation}
(The factor $\sqrt2$ in (\ref{r1}-\ref{r4}) comes from the relation
$\left\{\gamma^R,\gamma^L\right\}=2$, and the sign at it is the
matter of bispinor normalization convention.)

We shall nowhere need the use of matrix $\gamma_5$, for which the
chirality bispinors are eigenvectors. Thanks that all the momenta
are contained in one hyper-plane, one can manage with matrices
$\gamma^R$, $\gamma^L$ alone, playing the role of (nilpotent)
angular momentum raising and lowering operators.

Now, the smallest block in the matrix element
\begin{eqnarray}\label{gamma mu gamma mu}
    \gamma^\mu d_R\bar u'_R\gamma^\mu&=&-2d_L\bar u'_L,\\
    \gamma^\mu d_R\bar u'_L\gamma^\mu&=&0.
\end{eqnarray}
Eq.~(\ref{gamma mu gamma mu}) implies that fermion angular momentum
projection onto the collision axis must flip after the vector boson
exchange, and the spins of the opposing fermions must exactly
correlate. Physically, that is natural, since a vector boson emitted
by a $M_z=+\frac12$ fermion has $M_z=+1$, so after the vector boson
emission the fermion acquires $M_z=-\frac12$, and the opposite
fermion must initially have $M_z=-\frac12$ to be able to absorb the
$M_z=+1$ boson.

Hence,
\begin{equation}\label{N1}
    \mathcal N^{(1)}_{RR,RR}=\mathcal N^{(1)}_{LL,LL}=-2s,
\end{equation}
whereas all the other helicity amplitudes equal zero.

The next larger block
\begin{equation}\label{nu}
    \gamma^\nu\not q_1\gamma^\mu d_R\bar u'_R\gamma^\mu\not
    q_1\gamma^\nu=-2\gamma^\nu\not q_1 d_L\bar u'_L\not
    q_1\gamma^\nu=4\not q_1 d_R\bar u'_R\not
    q_1.
\end{equation}
The non-zero part of r. h. s. of (\ref{nu})
\begin{equation}\label{q1 q1}
    \not q_1d_R\bar u'_R\not q_1=\left(q_1^-\gamma^++q_1^R\gamma^L\right)d_R\bar
    u'_R\left(q_1^-\gamma^++q_1^L\gamma^R\right).
\end{equation}
Now, matrix-vectors $\not q_i$ sandwiching this expression have
components $q_i^+$ at $\gamma^-$, which are negligible as compared
to then $q_{i\perp}$. Then, it is possible to (anti-)commute the
matrices $\gamma^+$ in (\ref{q1 q1}) outwards to a position next to
on-mass-shell bispinors $\bar d'$ and $u$, action on which, by
virtue of (\ref{massless Dirac}), gives zero. So, block (\ref{nu})
equals
\begin{equation}\label{q1 q1}
    \not q_1d_R\bar u'_R\not q_1=q_1^Rq_1^L\gamma^Ld_R\bar
    u'_R\gamma^R=\vec q_\perp^2d_L\bar u'_L,
\end{equation}
which is proportional to (\ref{gamma mu gamma mu}).

Ultimately, it is understood that in the arbitrary order
\begin{equation}\label{N n}
    \mathcal N^{(n)}_{RR,RR}=\mathcal N^{(n)}_{LL,LL}=(-2)^ns\,\vec
    q_{1\perp}^2\ldots\vec
    q_{n-1\perp}^2.
\end{equation}

Note that the $2\vec q_\perp^2$ factors emerge here without the
appeal to the azimuthal averaging, or reasoning that $q_\parallel$
components cancel the logarithmic singularities in the integral (cf.
\cite{FGGL}). As is known, vector interaction at hard momentum
transfers (compared to the mass) is predominantly magnetic --
similarly to the conventional separation of electric and magnetic
form-factors:
\begin{equation}\label{FFs}
    J_{fi}^\mu=\bar
    u_f\left[F_e\left(Q^2\right)\gamma_\parallel^\mu+F_m\left(Q^2\right)\gamma_\perp^\mu\right]u_i.
\end{equation}

Since in our case polarizations of all the virtual particles are
completely fixed by that of initial ones, the problem is equivalent
to some scalar field theory. The vector character of the bosons does
not entail any momentum-dependent numerators, and merely secures
helicity conservation.

\subparagraph{\textbf{Color matrix factor.}}\label{color} As had
been discussed in \cite{FGGL}, \cite{Kirschner,KL}, in the perfect
charge (color) exchange situation the infra-red vector boson
exchange contributions mutually cancel. Here, let us neglect them
altogether, and consider only the hard ladder.

Embarking on the Fierz-type identity for color generators
\begin{equation}\label{Fierz color}
    t^A_{l'l}t^A_{m'm}=\frac12\delta_{l'm}\delta_{m'l}-\frac{1}{2N_c}\delta_{l'l}\delta_{m'm},
\end{equation}
by induction one proves\footnote{Here, $C_F$ and $-\frac{1}{2N_c}$
are just the values of Kirschner's matrix $\tau_2$ (defined in a
basis of convenience for him \cite{Kirschner-Yad.Fiz.}), and
$\delta_{m'l}\delta_{l'm}$ together with
$t^A_{m'l}t^A_{l'm}-\frac{1}{2N_c}\delta_{m'l}\delta_{l'm}$ are its
eigenvectors.}
\begin{eqnarray}\label{color-matr}
    \mathsf C^{(n)}&=&\left(t^{A_n}\ldots t^{A_1}\right)_{l'l}\left(t^{A_1}\ldots
    t^{A_n}\right)_{m'm}\nonumber\\
    &=&C_F^n\frac{1}{N_c}\delta_{m'l}\delta_{l'm}+2\left(-\frac{1}{2N_c}\right)^n\left[t^A_{m'l}t^A_{l'm}-\frac{1}{2N_c}\delta_{m'l}\delta_{l'm}\right],
\end{eqnarray}
with $C_F$ given by Eq.~(\ref{C_F}). Obviously,
$C_F>-\frac{1}{2N_c}$, both by sign, and in magnitude. At $n\geq2$,
i. e., in any loop, it suffices to keep only the first term in the
r.h.s. of (\ref{color-matr}). As the Kronecker symbols indicate, the
leading term requires exchange of color.\footnote{The second term of
(\ref{color-matr}), in fact, is not yet related to a self-consistent
scattering amplitude since it is devoid of infra-red DL
corrections.}

The underlying reason for the law that the color exchange is assured
at the given ordering of gluon emission and re-absorption, and when
$N_c\to\infty$, is also transparent. For each quark, the first
ladder gluon emitted by it carries away its color, and in addition
has arbitrary (except at the tree level) anticolor. The final quark
moving in the same direction will absorb this gluon last of all, and
must annihilate its anticolor whatever it is (by color
conservation), and accept its color. Thereby, the color of the final
quark will coincide with that of the comoving initial one.

Summarizing this section, re-absorption of gauge bosons in the
inverse order stipulates transfer of all the quantum numbers between
the scattered quarks. The large-$N_c$ limit here is sufficiently
robust, and within it the picture is equivalent to that of QED, the
coupling constant correspondence being $\alpha_{QED}\to\alpha_s
C_F$.

\section{Loop integrals in DLLA}\label{sec:denominators}
Using the numerator kinematical factors, we are in a position to
treat the loop integrals.

\subparagraph{\textbf{One-loop integral reduction. Wave-function
interpretation.}}

By far the simplest approach for of high-energy asymptotics
derivation and understanding is infinite momentum frame quantum
field theory. One might anticipate its applicability for the
backward scattering, as well, inasmuch as the denominator structure
in Feynman integrals is the same as for forward scattering. But,
because of the occurrence of factors $\vec q_\perp^2$ in the
numerator (see Sec.~\ref{sec:spin}), application of LCPT is
obstructed by the divergence of the eikonal integral, over
$d^2q_\perp$. To keep the treatment consistent, one may, first,
straightforwardly carry out the $q^-$ integration in Feynman
integrals. In one loop,
\begin{eqnarray}\label{}
    M^{(2)}/\mathsf C^{(2)}&\approx&-2is(4\pi\alpha)^2\int\frac{d^4q}{(2\pi)^4}\frac{2q_\perp^2}{(2p'_dq-i0)(2p_dq-i0)q^2(q-\Delta)^2}\nonumber\\
    &\approx&(4\pi\alpha)^2\int\frac{dq^-}{2\pi i(q^--i0)}\frac{dq^+}{2\pi
    q^+}\frac{d^2q_\perp}{(2\pi)^2}\frac{2q_\perp^2}{q_\perp^2(\Delta_\perp-q_\perp)^2}.\nonumber
\end{eqnarray}
Upon the integration (in the exact expression) over $q^-$, reducing,
essentially, to taking residue in a single pole $q^-\approx0$, we
derive a restriction on $q^+$: $p_d^-\leq q^+\leq p_d^+$. Then, one
can pass to the eikonal approximation. At that, the condition
\[
-(p'_d-q)^2\approx2p'_dq\approx2p'^-_dq^+\gg q^2,\, q_\perp^2
\]
yields ordering of $q_\perp^2$ and $q^+$, which secures convergence
of the integral over $q_\perp^2$ at large $q_\perp^2$. Within the
(double-) logarithmic accuracy,
\begin{eqnarray}
    M^{(2)}/\mathsf
    C^{(2)}&\approx&\frac{(4\pi\alpha)^2}{(2\pi)^2}\int_{p'^+_d}^{p_d^+}\frac{dq^+}{q^+}\int_{1}^{p'^-_dq^+}\frac{dq_\perp^2}{q_\perp^2}=\frac{(4\pi\alpha)^2}{8\pi^2}\ln^2s\label{M2/C2}\\
    &=&\left(M^{(1)}/\mathsf
    C^{(1)}\right)\frac\alpha{4\pi}\ln^2s.\nonumber
\end{eqnarray}

In the final representation (\ref{M2/C2}) valuable is the separation
of hard and soft physics, which does not in fact depend on our
choice of prior integration over $q^-$, or $q^+$. The longitudinal
hard gluons pertain to hard physics, whereas the braking fermions --
to the soft. Soft physics is most conveniently interpreted in terms
of wave functions and their overlaps. If one invokes the analogy
with the non-relativistic (or old-fashioned) perturbation theory,
Eq.~(\ref{M2/C2}) may be compared with the expression for the
second-order transition matrix element
\begin{equation}\label{non-rel 2}
    \left\langle2\right|V\left|1\right\rangle=\sum_n\frac{\left\langle2\right|V\left|n\right\rangle\left\langle
    n\right|V\left|1\right\rangle}{E_0-E_n}.
\end{equation}
The role of the perturbation operator $V$ in our case is played by
the coupling constant $4\pi\alpha$. The energy denominator finds an
analog in the factor $\frac1{q^+}$, which, however, is positive, not
negative, i. e., the intermediate states reside \emph{under} the
mass-shell. As for the intermediate state wave functions
$\left|n\right\rangle$, their counterparts are the factors
$\frac{\sqrt2}{q^1\pm iq^2}$. Finally, the phase space volume
element is $\frac{dq^+d^2q_\perp}{(2\pi)^3}$. It should be noted
that the phase space available for $q_\perp^2$ is restricted by the
value of the ``energy" $q^+$. That reflects the circumstance that
soft and hard physics are not separated absolutely, but only within
the logarithmic accuracy. A similar situation (not encountered at
forward scattering) is often met at description of exclusive
hadronic processes with a large momentum transfer (see, e. g.,
\cite{Lepage}).

In conclusion, let us remark that in \cite{FGGL} the extraction of
DLLA contributions is conducted by prior integration over
$d^2q_\perp$, in analogy with the Sudakov's vertex asymptotic
treatment \cite{Sudakov}. That renders the framework more symmetric
appearance, but the wave-function interpretation gets obscured.

\subparagraph{\textbf{All-order treatment.}} Integrals for higher
orders of perturbation theory may also be calculated via first
$q^-$-integration, but it requires more detailed considerations (cf.
\cite{ChengWu}). Instead of the variables $q_i^-$, $q_{i\perp}^2$,
in the present case it is convenient to introduce
\begin{equation}\label{}
    q_i^-, \quad \kappa_i=\frac{\vec q^2_{i\perp}}{2q_i^-},
\end{equation}
and only then carry out the $q_i^-$ integration. Then, a strong
ordering condition ensues
\begin{equation}\label{kappa ord}
    \kappa_i\gg\kappa_{i+1}
\end{equation}
(corresponding to the multiperipheral condition $q_i^-\ll q_{i+1}^-$
for $q_i^-$, which have been integrated over), and
\begin{equation}\label{kappa<q+}
    \kappa_i<q_i^+
\end{equation}
(corresponding to the eikonal condition $\vec
q^2_{i\perp}<2q_i^+q_i^-$). The integral of the 2n-th order of
perturbation theory, in DLLA assumes the form
\begin{eqnarray}\label{Mn/Cn}
    M^{(n)}/\mathsf
    C^{(n)}&=&\left(M^{(1)}/\mathsf
    C^{(1)}\right)\left(\frac\alpha{2\pi}\right)^{n-1}\int_{p'^+_d}^{p^+_d}\frac{dq_{1}^+}{q_{1}^+}\int_{p'^+_d}^{q_{1}^+}\frac{d\kappa_{1}}{\kappa_{1}}
    \ldots\nonumber\\
    &\!&\ldots\int_{p'^+_d}^{q_{n-3}^+}\frac{dq_{n-2}^+}{q_{n-2}^+}\int_{p'^+_d}^{\min(q_{n-2}^+,\kappa_{n-3})}\frac{d\kappa_{n-2}}{\kappa_{n-2}}
    \int_{p'^+_d}^{q_{n-2}^+}\frac{dq_{n-1}^+}{q_{n-1}^+}\int_{p'^+_d}^{\min(q_{n-1}^+,\kappa_{n-2})}\frac{d\kappa_{n-1}}{\kappa_{n-1}}.
\end{eqnarray}

For evaluation of this integral, it is convenient to recast the
$i$-th pair of integrations
\begin{eqnarray}\label{recast}
    \int_{p'^+_d}^{q_{i+1}^+}\frac{dq_i^+}{q_i^+}\int_{p'^+_d}^{\min(q_i^+,\kappa_{i+1})}\frac{d\kappa_i}{\kappa_i}\ldots
    &=&\int_{p'^+_d}^{q_{i+1}^+}\frac{dq_i^+}{q_i^+}\int_{p'^+_d}^{\kappa_{i+1}}\frac{d\kappa_i}{\kappa_i}\ldots
    -\int_{p'^+_d}^{\kappa_{i+1}}\frac{dq_i^+}{q_i^+}\int_{q_i^+}^{\kappa_{i+1}}\frac{d\kappa_i}{\kappa_i}\ldots\nonumber\\
    &\equiv&\int_{p'^+_d}^{q_{i+1}^+}\frac{dq_i^+}{q_i^+}\int_{p'^+_d}^{\kappa_{i+1}}\frac{d\kappa_i}{\kappa_i}\ldots
    -\int_{p'^+_d}^{\kappa_{i+1}}\frac{d\kappa_i}{\kappa_i}\int_{p'^+_d}^{\kappa_i}\frac{dq_i^+}{q_i^+}\ldots\nonumber\\
\end{eqnarray}
(the integral over a trapezium represented as an integral over the
rectangle minus the integral over the triangle). But, as is easy to
demonstrate by changing the order of variables,
\begin{equation}\label{}
    \int_{p'^+_d}^{\kappa_{i+1}}\frac{d\kappa_i}{\kappa_i}\int_{p'^+_d}^{\kappa_i}\frac{dq_i^+}{q_i^+}\left\{\int_{p'^+_d}^{q_i^+}\frac{dq_{i-1}^+}{q_{i-1}^+}\int_{p'^+_d}^{\kappa_i}\frac{d\kappa_i}{\kappa_{i-1}}
    -\int_{p'^+_d}^{\kappa_i}\frac{d\kappa_{i-1}}{\kappa_{i-1}}\int_{p'^+_d}^{\kappa_{i-1}}\frac{dq_{i-1}^+}{q_{i-1}^+}\right\}\ldots\equiv0,
\end{equation}
so, we can drop the terms
$-\int_{p'^+_d}^{\kappa_{i+1}}\frac{d\kappa_i}{\kappa_i}\int_{p'^+_d}^{\kappa_i}\frac{dq_i^+}{q_i^+}$
at all the $dq_i^+d\kappa_i$ integrations but the
$(n-1)$-th.\footnote{This means that inequalities (\ref{kappa<q+})
are inconsequential, except for the first and the last one.} The
$(n-1)$-th double integration gives
\begin{equation}\label{int1}
   \int_{p'^+_d}^{q_{n-2}^+}\frac{dq_{n-1}^+}{q_{n-1}^+}\int_{p'^+_d}^{\kappa_{n-2}}\frac{d\kappa_{n-1}}{\kappa_{n-1}}
    -\int_{p'^+_d}^{\kappa_{n-2}}\frac{d\kappa_{n-1}}{\kappa_{n-1}}\int_{p'^+_d}^{\kappa_{n-1}}\frac{dq_{n-1}^+}{q_{n-1}^+}=\ln\frac{q_{n-2}^+}{p'^+_d}\ln\frac{\kappa_{n-2}}{p'^+_d}-\frac12\ln^2\frac{\kappa_{n-2}}{p'^+_d}.
\end{equation}
Passing to the self-suggestive variables
\begin{equation}\label{}
    \eta_i=\ln\frac{q_i^+}{p'_d},\quad
    \xi_i=\ln\frac{\kappa_i}{p'_d},
\end{equation}
the DLLA amplitude of $2n$-th order is calculated quite trivially:
\begin{eqnarray}\label{}
    M^{(n)}/\mathsf
    C^{(n)}&=&\left(M^{(1)}/\mathsf
    C^{(1)}\right)\left(\frac\alpha{2\pi}\right)^{n-1}\int_0^{\ln
    s}d\eta_1\ldots\int_0^{\eta_{n-3}}d\eta_{n-2}\int_0^{\ln
    s}d\xi_1\ldots\int_0^{\xi{n-3}}d\xi_{n-2}\left\{\eta_{n-2}\xi_{n-2}-\frac12\xi^2_{n-2}\right\}\nonumber\\
    &=&\left(M^{(1)}/\mathsf
    C^{(1)}\right)\left(\frac\alpha{2\pi}\ln^2s\right)^{n-1}\left\{\frac{1}{[(n-1)!]^2}-\frac{1}{(n-2)!n!}\right\}\nonumber\\
    &=&\left(M^{(1)}/\mathsf
    C^{(1)}\right)\left(\frac\alpha{2\pi}\ln^2s\right)^{n-1}\frac1{(n-1)!n!}.
\end{eqnarray}
Invoking the series expansion for the modified Bessel function of
first order,
\begin{equation}\label{I1}
    I_1(z)=\sum_{n=1}^\infty\frac{\left(\frac
    z2\right)^{2n-1}}{(n-1)!n!},
\end{equation}
we arrive at the result of \cite{FGGL}, which we have thus
re-derived by straightforward resummation, without the recourse to
the formalism of integral equations.

Post factum, it is important to check the self-consistency of the
adopted multi-peripheral approximation (\ref{+ ordering}-\ref{q perp
ll q+q-}). When $z=\sqrt{\frac{2\alpha C_F}{\pi}}\ln s\gg1$, the
largest terms in sum (\ref{I1}) have numbers
\begin{equation}\label{n mean}
    \bar n\sim\frac{z}{2}.
\end{equation}
Equally, and independently of the overall energy, one can say that
each gluon typically shifts the quark rapidity by
\begin{equation}\label{}
    \Delta y=\frac{Y=\ln s}{\bar n}\sim\sqrt{\frac{2\pi}{\alpha
    C_F}}\simeq 5\div7.
\end{equation}
This implies that for the given problem the multi-peripheral
approximation is very safe.

The transverse motion of quarks in the ladder rails is usually
regarded as random walk. At that, the rung gluons propagate nearly
forward (since, in the eikonal approximation, their propagators do
not depend on transverse momenta), and so, impact parameters of the
final $u$-quark must coincide with that of the initial $d$-quark,
and impact parameter of the final $d$-quark -- with that of the
initial $u$-quark. Thereby, the walk is not completely random. Yet,
the walk step is small as compared to typical hadronic radius
(recall that large $\vec q_\perp^2$ dominate), so the initial quark
impact parameters must be close to one another, anyway.

\section{Discussion and summary}
The mechanistic picture of reggeization is observed to fall into
certain contrast with the analyticity and duality expectations. In
particular, transversal hardness excludes exact analogy with meson
exchange in the $t$-channel, and yet suggests that the
quark-exchange reggeization phenomenon, and the relevant intercept,
may be universal, or there can be a few universal reggeons (much
less numerous then the host of mesons). The similarity of the Regge
ladder diagram with that of the Bethe-Salpeter equation must not be
deluding, given the dominance in the present case of large $q_\perp$
(let alone the excessive hardness of the ladder $u$-channel gluons).
In their own turn, mesons, being strongly bound relativistic states,
for which the interaction radius is not small compared to the
average inter-constituent distance must not necessarily obey a
Bethe-Salpeter-like equation at all.

In what concerns the hadron wave function overlap representation,
the hardness of the Regge ladder implies that one can rather safely
exploit the notion of coincidence of colliding quark impact
parameters -- unless the energy becomes super-high, giving the
short-step transverse random walk eventually a spread comparable to
the hadron size. Another feature important at hadron wave function
overlap computations is that the reggeized kernel (\ref{regge th})
is not scale-invariant, and does not factorize in terms of Feynman-x
of the active quarks:
\begin{equation}\label{}
    M_{qq'\to q'q}\left(s_{qq'}=s_{hh}x_qx_{q'}\right)\neq
    f_1\left(s_{hh}\right)f_2\left(x_q\right)f_3\left(x_{q'}\right),
\end{equation}
and it is only in far asymptotics (\ref{regge asympt}), where some
noninteger-power scaling law and factorization set in. Finally, note
that amplitude (\ref{regge th}) is neither even, nor odd function of
$s$, in contrast to the kernel in Born approximation. The latter
property matters at calculations of meson flavor exchange
amplitudes.

\end{document}